\documentstyle[12pt]{article}  
  
\def\12{\frac{1}{2}}
\def\14{\frac{1}{4}}

\def\g{\gamma}

\newcommand{\be}{\begin{equation}} 
\newcommand{\ee}{ \end{equation}}
\newcommand{\ba}{\begin{array}}
\newcommand{\ea}{\end{array}}
\newcommand{\bea}{\begin{eqnarray}}
\newcommand{\eea}{\end{eqnarray}}

\begin{document}  
  
\begin{titlepage}

\begin{flushright}  
 
HUB-EP-00/005\\
   
\end{flushright}  
  
\vspace{3cm}  
\begin{center}  
  
{\Large \bf{Actions for Curved Branes}}  
  
\vspace{1cm}  
  
Mohab Abou-Zeid\footnote{abouzeid@physik.hu-berlin.de} 

\vspace{.3cm}  
{\em Institut f\"{u}r Physik, Humboldt Universit\"{a}t zu Berlin, 
Invalidenstrasse 110, \\ D-10115   Berlin, Germany}   

\vspace{2cm}  
\begin{abstract}
The nondeterminantal forms of the Born-Infeld and related brane actions in 
which the gauge fields couple to both an induced metric and an intrinsic 
metric are generalised by letting either or both metrics be dynamical. The 
resulting actions describe  \lq 
brane world' and cosmological scenarios in which the gauge fields are 
confined to the brane, while gravity propagates in both the world-volume and 
the bulk. In particular, for actions involving a nonsymmetric \lq metric', 
nonsymmetric gravity propagates on the worldvolume. For 3-branes with a 
symmetric metric, conformal (Weyl) gravity propagates on the worldvolume and 
has conformally invariant couplings to the gauge fields.
\end{abstract}
  
\end{center}  
\vspace{3cm}
 
\flushleft{January, 2000}
 
\end{titlepage}


\section{Introduction}

Recently, there has been considerable interest in cosmological and elementary 
particle physics models in which the usual particles and interactions of the 
Standard Model are confined to an effectively $(3+1)$-dimensional 
hypersurface embedded in  a higher-dimensional spacetime~\cite{RSr,RSs}. In 
such models, gravity is allowed to propagate 
in the bulk of the dimensions transverse to the hypersurface, so that the 
hypersurface must also be dynamical. Although there are various ways of 
implementing such scenarios, including those in which the gravitational and 
gauge interactions are unified at 
the electroweak scale and there are large compact dimensions~\cite{ADD}, 
string theory provides attractive constructions via 
$D$-branes~\cite{AADD,ST,KT}. An apparently different scenario was proposed 
in~\cite{RS}, where it is argued that the so-called 
warped compactifications provide phenomenologically attractive alternatives 
to compactification manifolds. In the construction of~\cite{RS},  space-time 
is taken to be an $AdS_5$ 
space with the near-boundary region cut away and replaced with a wall of 
constant extrinsic curvature. It is then found that there is a normalizable 
graviton zero  mode in the four dimensional boundary.  In~\cite{SSG} (and 
references quoted therein), it 
has been suggested  on the basis of the AdS/CFT 
correspondence~\cite{JM,GKP,EW} that this construction amounts to a coupling 
of gravity to a certain boundary conformal field theory. However, it remains
to be seen whether this setup can be embedded in string theory.

In the present paper, we will propose an alternative approach to the 
construction of brane world models with dynamical worldvolume gravity 
interacting with gauge and other matter fields. This utilizes the 
nondeterminantal forms of the Born-Infeld and related Dirac-Born-Infeld 
actions for $p$-branes found in 
refs.~\cite{AH1,AH2}. In the particular case of the 3-brane, the action 
found in~\cite{AH2} involves an intrinsic metric and has a Weyl invariance 
under rescalings of the metric. Since there are several reasons for imposing 
scale invariance on the brane world action~\cite{Adler}, we seek a coupling 
to worldvolume gravity which preserves this symmetry at the classical 
level. This is achieved by promoting the intrinsic metric to a dynamical 
field with dynamics determined by scale-invariant Weyl gravity, and is 
similar in spirit to an earlier proposal based on the Weyl-Dirac-Yang-Mills 
action~\cite{Zee1,Zee2}. The quantum theory of the conformally invariant Weyl 
action has a number of remarkable features which are desirable in a 
fundamental theory for quantum gravity. For example, it is known to reduce to 
Einstein gravity in the low energy limit~\cite{Adler} and there are 
indications that it may be renormalizable~\cite{Stelle,Zee1} and 
asymptotically free~\cite{FT}. It is intriguing that a nonconformal theory 
such as Born-Infeld theory can in fact be coupled to gravity in a conformally 
invariant manner.

\section{Born-Infeld and Brane Actions}

       The Nambu-Goto action for a $p$-brane with $p=n-1$ is
\begin{equation}
S_{NG} = -T_p \int d^n \sigma \sqrt{-\det{ \left( G_{\mu \nu} 
\right) }}   ,    \label{NG}
\end{equation}
where   $T_p$ is the $p$-brane
tension and
\begin{equation}
G_{\mu \nu} = G_{ij} \partial_\mu X^i \partial_\nu X^j   \label{induced}
\end{equation}
is the world-volume metric induced  by the   spacetime metric $G_{ij}$.
The non-linear form of the action~(\ref{NG}) is inconvenient
for many purposes, including that of quantisation. However, introducing an 
intrinsic  worldvolume metric $
g_{\mu
\nu}$ allows one to write down the equivalent action~\cite{Polya,BVH,HT}
\begin{equation}
S_P = -\frac{1}{2}T'_p \int d^n \sigma \sqrt{-g} \left[ g^{\mu \nu} 
G_{\mu \nu} -(n-2)
\Lambda \right] ,
\label{Polyakov}
\end{equation}
where   $g \equiv \det{(g_{\mu
\nu})}$ and $\Lambda$ is a   constant. The metric $g_{\mu \nu}$ is an auxiliary
field which can be eliminated using its equation of motion
to recover  action~(\ref{NG}). The  constants $T_p$ 
and $T'_{p}$ are related by 
\begin{equation}
T'_p =    \Lambda^{\frac{n}{2}-1} T_p .
\label{tensions}
\end{equation}
This form of the action is much more convenient than~(\ref{NG}), as it is
quadratic in $\partial X$. In particular, for strings ($n=2$), 
Polyakov~\cite{Polya} showed that the functional integral over all closed 
compact worldsheets can be reduced to the quantum theory of the 
two-dimensional Liouville Lagrangian. However, the explicit evaluation of the 
functional integrals~\cite{Polya} relied on the fact, 
special to two dimensions, that it is possible to change coordinates in such 
a way that the worldsheet metric becomes conformally Euclidean. Thus the 
integration over $X$ yields
 an integral which depends on the conformal factor only through the conformal 
anomaly.

        The Born-Infeld action for a
 vector field $A_\mu$ in
an $n$-dimensional space-time with metric $G_{\mu\nu}$ is
\begin{equation}
S_{BI} = -T_p \int d^n \sigma \sqrt{-\det{\left( G_{\mu \nu} +F_{\mu \nu}
\right) }} ,
\label{BI}
\end{equation}
where   $F=dA$ is the Maxwell field
strength.
A related $(n-1)$-brane action is
 \begin{equation}
S_{DBI} = -T_p \int d^n \sigma \sqrt{-\det{\left( G_{\mu \nu} +
\cal{F}_{\mu \nu} \right) }} ,
\label{DBIB}
\end{equation}
where $G_{\mu \nu}$ is the induced metric~(\ref{induced}) and
${\cal F}_{\mu \nu}$ is the antisymmetric tensor field
\begin{equation}
{\cal{F}}_{\mu \nu} \equiv F_{\mu \nu} -B_{\mu \nu}      \label{defF}
\end{equation}
with $B_{\mu\nu} $ the pull-back of a space-time 2-form gauge field $B$,
\begin{equation}
B_{\mu \nu} = B_{ij} \partial_\mu X^i \partial_\nu X^j   \label{defB} .
\end{equation}
The action~(\ref{DBIB}) is closely related to the D-brane 
action~\cite{FTBI,RGL,EWp,CS}
 and the Born-Infeld action~(\ref{BI}) can be thought of
as a special case of this, but with a different interpretation of $G_{\mu\nu}$.
Just as in the case of the action~(\ref{NG}), the non-linearity 
of~(\ref{DBIB})  makes
it rather difficult to study. However, a nondeterminantal  action which is 
the analog of~(\ref{Polyakov}) 
for this case has been proposed in~\cite{AH1}, based on the introduction of  
an auxiliary
world-volume tensor
  field $\g_{\mu \nu}$ with both a symmetric part $\g_{(\mu \nu )}$  and an 
antisymmetric part $\g_{[\mu \nu ]}$. Such \lq metrics' have been  used in 
alternative theories of gravitation; see e.~g.~\cite{AE,JWM1}. The action 
which is classically equivalent
 to~(\ref{DBIB}) is
\begin{equation}
S' = - \frac{1}{2}T'_p \int d^n \sigma \sqrt{-\g} \left[ ( {\g}^{-1})^{\mu 
\nu } \left( G_{\mu \nu} +
\cal{F}_{\mu \nu} \right)
- (n-2)\Lambda \right] ,
\label{EH}
\end{equation}
where $\g \equiv \det{\left( \g_{\mu \nu}\right)}$; the inverse tensor 
$({\g}^{-1})^{\mu \nu}$ satisfies
\begin{equation}
({\g}^{-1} )^{\mu \nu} \g_{\nu \rho} = \delta^{\mu}{}_{\rho} \label{defginv} .
\end{equation}
For $n \ne 2$, the    $\g_{\mu \nu}$ field equation implies   
\begin{equation}
  G_{\mu \nu} +
{\cal{F}}_{\mu \nu} = \Lambda \g_{\nu \mu}       \label{geom}
\end{equation}
and substituting  back into~(\ref{EH}) yields the Born-Infeld-type 
action~(\ref{DBIB})
where the constants $T_p,T'_p$ are related as in eq.~(\ref{tensions}). 
For $n=2$, the action~(\ref{EH}) is invariant under the generalised Weyl
transformation
\begin{equation}
\g_{\mu\nu} \to \omega (\sigma) \g_{\mu\nu} 
\label{symm}
\end{equation}
and the $\g_{\mu \nu}$ field equation implies   
\begin{equation}
 G_{\mu \nu} + {\cal F}_{\mu \nu}  =  \Omega \g_{\nu \mu}       \label{geom2}
\end{equation}
for some conformal factor $\Omega$.

The action~(\ref{EH}) is {\em linear} in $ \left( G_{\mu \nu} +
\cal{F}_{\mu \nu} \right)
$ and so is much easier to analyse
than (\ref{DBIB}).
In particular, it is linear in $F$ and quadratic in $\partial X$.

As was shown in~\cite{AH2}, the action~(\ref{DBIB}) can also be rewritten in a 
form which is quadratic in the
field strength $F$, and is therefore simpler to analyse and quantise. This 
form of the action uses an intrinsic worldvolume metric $g_{\mu \nu}$ (for 
the remaining part of this section, we write $g_{\mu\nu} \equiv 
\g_{(\mu\nu )}$ with $\g_{[\mu \nu]}=0$), just 
like~(\ref{Polyakov}). The action which is classically equivalent 
to~(\ref{DBIB}) and  is quadratic in the gauge  
field
strength  $F_{\mu \nu}$ is
\begin{equation}
S''  = -T''_p \int
d^{p+1}\sigma (-G)^{\14} (-g )^{\14} \left[ g^{\mu \nu}
\left( G_{\mu \nu} -G^{\rho \sigma} \cal{F}_{\mu \rho} \cal{F}_{\sigma \nu} 
\right)
- (p-3) \Lambda \right] ,
\label{new1}
\end{equation}
where $g \equiv \det (g_{\mu \nu})$ and $\Lambda$ is a constant.
For
$p\neq 3$, the $g_{\mu \nu}$ field equation implies
\begin{equation}
g_{\mu \nu} = \frac{1}{\Lambda} \left( G_{\mu \nu} -G^{\rho \sigma}
\cal{F}_{\mu \rho} \cal{F}_{\sigma \nu} \right)
\label{fenot3}
\end{equation}
and substituting back into~(\ref{new1}) yields action~(\ref{DBIB}). The 
constants $T_p$,
$T_p ''$ are
related by
\begin{equation}
T_p '' =\14 \Lambda^{\frac{p-3}{4}} T_p .
\label{T's}
\end{equation}

For $p=3$, the  four-dimensional action~(\ref{new1}) is invariant under the
usual Weyl transformation
\begin{equation}
g_{\mu \nu} \rightarrow \omega (\sigma ) g_{\mu \nu}
\label{Weyl}
\end{equation}
and the $g_{\mu \nu}$ field equation implies
\begin{equation}
g_{\mu \nu} = \Omega \left( G_{\mu \nu} -G^{\rho \sigma}
\cal{F}_{\mu \rho} \cal{F}_{\sigma \nu} \right)
\label{gfe3}
\end{equation}
for some conformal factor $\Omega$.

The actions~(\ref{EH}) and~(\ref{new1}) are easily generalised to the 
D$p$-brane kinetic term
\begin{equation}
S=-T_p \int d^{p+1} \sigma e^{-\phi} \sqrt{ -\det ( G_{\mu \nu}
+{\cal F}_{\mu \nu} )}
\label{DBI}
\end{equation}
where $\phi$, $G_{\mu \nu}$ and $B_{\mu \nu}$ are the pullbacks to the
worldvolume of the background dilaton, background metric and background
NS antisymmetric two-form
fields,
$F=dA$, with $A$
the $U(1)$ world-volume gauge field, and $\cal{F}_{\mu\nu}$ is defined as 
in~(\ref{defF}) . This action gives the
effective dynamics of the zero-modes of the open strings with ends tethered
on a D-brane when
$F$ is slowly varying, so that corrections involving $\nabla F$ can be  
ignored,
and has
therefore played a central role in recent studies of
D-brane dynamics and string theory duality~\cite{JP}.

As shown in~\cite{AH1,AH2}, the methods above can also be applied to the
kappa-symmetric action for a D-brane~\cite{APS,CGNSW,BTD}, to the low 
energy effective action for an open  type I string~\cite{Arkady}, to 
static-gauge D$p$-brane actions~\cite{JHS},  and to the M-theory five-brane
action~\cite{PST,J5}. For example, a classically equivalent  form of the 
latter which is quadratic in the field strength of the worldvolume self-dual 
two-form tensor gauge field was given in~\cite{AH2} using a (symmetric) 
auxiliary worldvolume metric.

\section{Brane Effective Actions}

Whatever the nature of the fundamental theory at short distances, the  
space-time  effective action at large distances in the presence of a 
brane (which can be a D-brane, an M-brane or a more general domain wall) is of the form
\begin{equation}
S=S_{bulk}+S_{brane} .
\end{equation}
Here the action
\begin{equation}
S_{bulk} = \int d^D X (-G)^\frac{1}{2} \left[ -\Lambda +2M^{D-2}R(G) +\ldots 
\right] ,
\label{bulk}
\end{equation}
where $M$ is the $D$-dimensional Planck mass and $\Lambda$ the cosmological 
constant, describes the bulk gravitational and other massless fields. In 
string theory, $S_{bulk}$ is the bosonic part of the appropriate low-energy 
supergravity theory, e.~g.\ type IIA or type IIB supergravity in the 
Einstein frame; in M-theory, it is the bosonic part of eleven dimensional 
supergravity~\cite{CJS}. In both cases, $S_{bulk}$ receives higher curvature 
corrections and also includes other terms for the remaining massles spacetime
fields; this is 
indicated by the ellipsis in~(\ref{bulk}). The action $S_{brane}$ describes 
the effective dynamics of the 
massless bosonic modes on the brane worldvolume. For example,  the bosonic 
part of the
effective world-volume action for a D-brane in a type II supergravity 
background contains the terms~\cite{MD1,GHT,GHM}
\begin{equation}
S_{Dbrane} = -T_p \int d^n \sigma e^{-\phi} \sqrt{-\det{\left( G_{\mu \nu} +
{\cal F}_{\mu \nu}
\right)}} +T_p \int_{W_{n}} C e^{\cal F} \left( 
\frac{\hat{\cal A}(R_T )}{\hat{\cal A} (R_N )} \right)^{\frac{1}{2}} .
\label{BI+WZ}
\end{equation}
The second term is a Wess-Zumino term and gives the coupling to the
background Ramond-Ramond
$r$-form gauge fields $C^{(r)}$ (where $r$ is odd for type IIA and even for 
type IIB) as well as the gravitational curvature effects induced from the 
bulk geometry in which the brane is embedded. The
potentials $C^{(r)}$ for $r>4$ are the duals of the potentials $C^{(8-r)}$.
 In~(\ref{BI+WZ}),
$C$ is the formal sum~\cite{GHT}
\begin{equation}
C \equiv \sum_{r=0}^{9} C^{(r)}  ,       \label{defC}
\end{equation}
all forms in space-time are pulled back to the worldvolume of the brane
$W_n$ and it is understood that the $n$-form part of $C e^{\cal F}$, which is 
$C^{(n)} +C^{(n-2)}
{\cal F} +\frac{1}{2} C^{(n-4)} {\cal F}^2 +\ldots $, is 
selected. In~(\ref{BI+WZ}), $\hat{\cal A}$ denotes the Dirac \lq roof' genus 
whose square root has an expansion in powers of the curvature two-form, and 
the components of the curvature are split: $R_T$ denotes components with 
tangent-space indices, while $R_N$ denotes components 
in the normal bundle (see~\cite{GHM} for notational details). 

The first term in~(\ref{BI+WZ}) receives corrections involving derivatives of 
the field
strength $F$~\cite{Arkady}, as well as gravitational curvature contributions
induced by the background geometry or by non-trivial worldvolume 
embeddings~\cite{BBG}. 

Introducing $\g_{\mu \nu}$ as before, we obtain the classically equivalent     
 D-brane action 
\begin{eqnarray}
S''_{Dbrane} &  = & -T''_p \int
d^{p+1}\sigma (-G)^{\14} (-g )^{\14}  e^{-\phi } \left[ g^{\mu \nu}
\left( G_{\mu \nu} -G^{\rho \sigma} \cal{F}_{\mu \rho} \cal{F}_{\sigma \nu} 
\right)
- (p-3) \Lambda \right]
\nonumber \\ & & + T_p \int_{W_{n}} Ce^{\cal F} \left( \frac{\hat{\cal A}
(R_T )}{\hat{\cal A} (R_N )} \right)^{\frac{1}{2}}.
\label{P+WZ}
\end{eqnarray}
 The field 
equation for $g_{\mu \nu}$ is given in~(\ref{fenot3}) 
and~(\ref{gfe3}); substituting
back into~(\ref{P+WZ}) yields~(\ref{BI+WZ}).

\section{Dynamical Gravity on the Brane}

In previous sections, the  worldvolume  tensor fields $\g_{\mu\nu}$
were merely auxiliary, 
i.~e.\ they were not dynamical. Motivated by the remarks in the introduction, 
we now propose to promote such fields to dynamical gravitational \lq metric' 
potentials on the world
volume. The new actions presented below are also of some interest in their 
own right, as they describe the dynamics of nonlinear gauge fields 
interacting non-trivially with gravity. In certain limits, the worldvolume 
dynamics of the gauge fields is of Born-Infeld type, while the dynamics  of 
the intrinsic \lq metric' $\g_{\mu\nu}$ is governed by Einstein gravity or an 
appropriate (symmetric) scale-invariant or a nonsymmetric generalisation 
thereof. In the 
case where the 
worldvolume tensor introduced is a symmetric metric $g_{\mu\nu}$, we also
 propose an  action which specifies  the dynamics of gauge fields coupled to 
{\em two} worldvolume metrics, namely $g_{\mu\nu}$ and the induced metric 
$G_{\mu\nu}$. In the particular case of the 3-brane, the Weyl invariance with 
respect to rescalings of $g_{\mu
\nu}$ suggests that the dynamics of this metric should be governed by a 
scale-invariant Weyl term.

We begin with the case of a symmetric intrinsic metric $g_{\mu\nu}$. We first 
consider dynamical induced gravity on the worldvolume, i.~e.\ we add an
 Einstein-Hilbert term as well as cosmological and higher derivative 
terms to~(\ref{new1}). Thus we  consider the action
\begin{eqnarray}
S''_{brane}  & = & -T''_p \int
d^{p+1}\sigma (-G)^{\14} (-g )^{\14} \left[ g^{\mu \nu}
\left( G_{\mu \nu} -G^{\rho \sigma} \cal{F}_{\mu \rho} \cal{F}_{\sigma \nu} 
\right)
- -(p-3) \Lambda_1 \right] \nonumber \\ & & +\lambda \int
d^{p+1}\sigma  (-G )^{\frac{1}{2}} \left[ \Lambda_2 + a R(G) + bR^2 (G) +
\ldots \right] .
\label{dyn1}
\end{eqnarray}
Here $R(G)$ denotes the scalar curvature of the induced metric $G_{\mu\nu}$
 and $\lambda$ is some coupling. Note that we have allowed the possibility of 
having a cosmological constant on the world-volume which differs from that of 
the bulk theory. This type 
of action does actually arise in string theory. For example, the $R^2 (G)$ 
and other  quadratic in curvature corrections to the Dirac-Born-Infeld part 
of the effective action for a single D$p$-brane were obtained in~\cite{BBG} 
to lowest order in the open
 string loop expansion. 

In~(\ref{dyn1}), the intrinsic metric is auxiliary and the worldvolume 
gauge fields have 
Born-Infeld-type  dynamics. Consider instead a dynamical intrinsic metric 
$g_{\mu\nu}$. Adding an Einstein-Hilbert and a   cosmological term 
to~(\ref{new1}) gives
\begin{eqnarray}
S''_{brane} & = & -T''_p \int
d^{p+1}\sigma (-G)^{\14} (-g )^{\14} \left[ g^{\mu \nu}
\left( G_{\mu \nu} -G^{\rho \sigma} F_{\mu \rho} F_{\sigma \nu} \right)
-(p-3) \Lambda_1 \right] \nonumber \\ & & +\lambda \int
d^{p+1}\sigma  (-g )^{\frac{1}{2}} \left[ \Lambda_2 + a R(g)  \right]
\label{dyn2}
\end{eqnarray}
where $R(g)$ denotes the scalar curvature of the intrinsic metric 
$g_{\mu\nu}$ and $\lambda$ is some coupling. Moreover, quadratic or higher 
curvature corrections $X_{\mu\nu}$ can be introduced via a coupling 
$\sqrt{-g}g^{\mu\nu}X_{\mu\nu}$. Although this
 is more natural in the present nondeterminantal framework, it is also 
possible to replace the second line in~(\ref{dyn2}) with a determinant 
gravity action of the form $\int d^n \sigma \det [ag_{\mu\nu}+bR_{\mu\nu} +
cX_{\mu\nu}]$; see~\cite{DG} for a discussion of physical requirements on 
such actions.

The case of a 3-brane is rather special as a result of the Weyl 
invariance~(\ref{Weyl}). Clearly this symmetry will not be preserved in the 
model described by~(\ref{dyn2}), because Einstein gravity is not Weyl 
invariant\footnote{Einstein gravity can be made locally Weyl invariant using 
a scalar compensating 
field~\cite{Deser}. This procedure can be adapted to make a certain
modification of 
Born-Infeld theory Weyl invariant, as discussed in~\cite{DG}.}. However it is 
possible to couple~(\ref{DBIB}) to gravity in a 
Weyl invariant manner by adding a (fourth derivative) Weyl term,
\begin{eqnarray}
S''_{brane}  & = & -T''_3 \int
d^{p+1}\sigma (-G)^{\14} (-g )^{\14}  g^{\mu \nu}
\left( G_{\mu \nu} -G^{\rho \sigma} \cal{F}_{\mu \rho} 
\cal{F}_{\sigma \nu} \right)
 \nonumber \\ & & +\lambda \int
d^{p+1}\sigma  (-g )^{\frac{1}{2}} C^{\mu}{}_{\nu\rho\sigma}(g) 
C^{\nu}{}_{\mu\lambda\tau}(g) g^{\lambda \rho}g^{\tau\sigma} 
\label{dyn3}
\end{eqnarray}
where $C^{\mu}{}_{\nu\lambda\rho}$ denotes the four-dimensional Weyl 
conformal tensor of $g_{\mu\nu}$,
\begin{eqnarray}
C^{\mu}{}_{\nu\lambda\rho} & \equiv & R^{\mu}{}_{\nu\lambda\rho}-\frac{1}{2} 
\left( \delta^{\mu}_{\lambda} R_{\nu\rho} - \delta^{\mu}_{\rho} 
R_{\nu\lambda} -g_{\nu\lambda} R^{\mu}{}_{\rho} +g_{\nu\rho} 
R^{\mu}{}_{\lambda} \right) \nonumber \\ & & + \frac{
1}{6} R \left( \delta^{\mu}_{\nu}g_{\lambda\rho} -\delta^{\mu}_{\rho} 
g_{\nu\lambda} \right) .
\end{eqnarray}
By construction, this is invariant under the transformation~(\ref{Weyl}).

As explained above, we can also couple the gauge fields covariantly  to both 
the intrinsic metric $g_{\mu\nu}$ and the induced metric $G_{\mu \nu}$ on the
$p$-brane via 
the action
\begin{eqnarray}
S''_{brane}  & = & -T''_p \int
d^{p+1}\sigma (-G)^{\14} (-g )^{\14} \left[ g^{\mu \nu}
\left( G_{\mu \nu} -G^{\rho \sigma} \cal{F}_{\mu \rho} 
\cal{F}_{\sigma \nu} \right)
 -(p-3) \Lambda_1 \right] \nonumber \\ & & +\lambda \int
d^{p+1}\sigma  (-g )^{\14}  (-G )^{\14}\left[ \Lambda_2 + R(g) R(G) \right]
\label{dyn4}
\end{eqnarray}
where $R(G)$ is the scalar curvature of the induced metric 
$G_{\mu\nu}$. Observe that the factors of the determinants ensure that the 
second line also transforms as a scalar under worldvolume 
diffeomorphisms.

Now consider the action~(\ref{EH}), in which the intrinsic \lq metric' 
$\g_{\mu\nu}$ has both a symmetric part $g_{\mu\nu }$ 
and an an 
antisymmetric part $\g_{[\mu\nu ]}$. We can promote this wordvolume tensor 
field to  dynamical gravitational fields by adding  terms which have 
previously been studied in the context of nonsymmetric 
gravitational theories, e.~g.\ in~\cite{JWM1}. Thus we consider the action  
\begin{eqnarray}
S'_{brane} & = & - \frac{1}{2}T'_p \int d^n \sigma \sqrt{-\g} \left[ 
( {\g}^{-1})^{\mu \nu } \left( G_{\mu \nu} +
\cal{F}_{\mu \nu} \right)
 -(n-2)\Lambda_1 \right] \nonumber \\ & & + \lambda \int d^n \sigma  L
\label{EHgrav}
\end{eqnarray}
where $\omega$ is an arbitrary coupling, and the Lagrangian density of the
non-Riemannian geometry is given by
\begin{equation}
L= \sqrt{-\g}\g^{\mu\nu} R_{\mu \nu}(W) -2\Lambda_1 \sqrt{-\g} -\14 \Lambda_2 
\sqrt{-\g}\g^{\mu\nu}
\g_{[\nu\mu ]}- \frac{1}{6}\g^{\mu\nu}W_\mu W_\nu
\end{equation}
Note the presence of the second cosmological constant $\Lambda_2$ associated 
with $\g_{[\mu \nu]}$. The curvature tensor $R_{\mu \nu}(W)$ is defined in 
terms
of unconstrained nonsymmetric connections
\begin{eqnarray}
W^{\lambda}_{\mu\nu} & = &\Gamma^{\lambda}_{\mu\nu} -\frac{2}{3} 
\delta^{\lambda}_{\mu} W_\nu \\ W_\mu & = & W^{\lambda}_{[\mu\lambda ]}
\end{eqnarray}
via the expression
\begin{equation}
R_{\mu\nu}(W)= \partial_\rho W^{\rho}_{\mu \nu} -\frac{1}{2} \left( 
\partial_\nu W^{\rho}_{\mu \rho} + \partial_\mu W^{\rho}_{\nu\rho} 
\right) -W^{\rho}_{\sigma\nu}W^{\sigma}_{\mu\rho} 
+W^{\rho}_{\sigma\rho}W^{\sigma}_{\mu\nu}
\end{equation}

The field equations for $\g_{\mu\nu}$, $G_{\mu\nu}$ and $F_{\mu\nu}$ can
be straightforwardly derived from the actions~(\ref{dyn1}), (\ref{dyn2}),
(\ref{dyn3}), (\ref{dyn4}) and~(\ref{EHgrav}). We expect these systems
of equations to 
have a rich set of interesting 
solutions, including worldvolume instantons, black holes, extended solitons 
and brane world cosmologies.

\section{Discussion}
  
In this paper, motivated by recent ideas on brane world  models and 
especially warped compactification  and their relation to the AdS/CFT 
correspondence,  we have 
considered various generalisations of the nondeterminantal forms of the 
Born-Infeld and related brane actions found in~\cite{AH1,AH2} which 
involve dynamical gravity on the brane worldvolume. In such
actions, the gauge fields couple both to a background or induced metric 
$G_{\mu\nu}$ and to an intrinsic tensor $\g_{\mu\nu}$. The latter can either 
be 
taken to be  symmetric, in which case the brane kinetic terms are quadratic in 
the  field strength $F$ of the abelian gauge fields, or to have both a 
symmetric and an antisymmetric part, in which case the brane kinetic terms are
linear in $F$. Either $G_{\mu\nu}$ or $\g_{\mu\nu}$ can be promoted to 
dynamical gravitational fields on the brane worldvolume by adding suitable 
gravitational contributions to the brane kinetic terms. Dynamical induced 
gravity occurs in string theory with D-branes, as seen e.~g.\ in 
refs.~\cite{GHM,BBG}. Dynamical intrinsic gravity does not occur in this
context, because the spectra of zero modes of the open strings tethered
on D-branes do not include gravitons. However, it  may turn out to 
have applications to brane world models in which 
a conformal or nonconformal quantum field theory on the brane is coupled to 
gravitational modes confined to the brane, as in~\cite{RS}. 

If a general intrinsic tensor
$\g_{\mu\nu}$ is chosen, then it is natural to let its worldvolume dynamics 
be given by the nonsymmetric gravitational theory~\cite{JWM1}. This leads to 
interesting brane world versions of the cosmological scenarios discussed 
in~\cite{JWM2}. For example, if the skew part of $\g_{\mu\nu}$ is assumed to
be small, then the nonsymmetric gravitational field equations can be expanded 
about the Friedmann-Robertson-Walker metric and the equation for the 
brane cosmological factor determines corrections due to a non-zero skew part
$\g_{[\mu\nu ]}$ and its derivatives~\cite{JWM2}. It would be interesting to
study such brane world models further.

If a symmetric intrinsic metric $g_{\mu\nu}$ is used instead, then it is more
natural to consider  Einstein 
gravity or a higher derivative symmetric gravitational  theory on the brane. In
the special case of 3-branes, the preservation of the generalised classical
Weyl invariance~(\ref{Weyl}) suggests that worldvolume gravity should be of
the Weyl type, as in action~(\ref{dyn3}). This describes the classical
scale-invariant dynamics  of  a fluctuating brane
world in which nonlinear abelian gauge fields couple to Weyl gravity. Quantum 
mechanically, the Weyl invariance~(\ref{Weyl}) will be anomalous, both in the
term quadratic in $F$ and in the $C^2$ Weyl 
term~\cite{FT83}. It would be very interesting to compute the conformal
anomaly for brane world models based on this action. Since~(\ref{dyn3}) can 
easily be generalised to include an
arbitrary number of abelian vectors and scalars conformally coupled to
Weyl gravity, it may be possible to cancel
the conformal anomaly. This would generalise the critical dimension
of string theory in the formulation of Polyakov, and may yield a consistent 
theory of dynamical 3-branes. Similar ideas were recently proposed
in~\cite{CS99} on the basis of the AdS/CFT correspondence.

\section*{Acknowledgements}

I would like  to thank Christopher Hull and Dieter L\"{u}st for helpful 
comments, and  the Theory 
Division at CERN for hospitality during the completion of this work. This 
research was supported by the Swiss National Science Foundation under grant 
number TMR83EU-056178.


\begin{thebibliography}{99}  

\bibitem{RSr} V.\ A.\ Rubakov and M.\ E.\ Shaposhnikov, {\em Do We Live 
Inside a Domain Wall?}, Phys.\ Lett.\ {\bf B125} (1983) 136.

\bibitem{RSs} R.\ Sundrum, {\em Effective Field Theory for a Three-Brane 
Universe}, Phys.\ Rev.\ {\bf D59} (1999) 085009, [hep-ph/9805471].

\bibitem{ADD} N.\ Arkami-Hamed, S.\ Dimopoulos and G.\ Dvali, {\em The 
Hierarchy Problem and New Dimensions at a Millimeter}, Phys.\ Lett.\ 
{\bf B429} (1998) 263, [hep-ph/9803315].

\bibitem{AADD} I.\ Antoniadis, N.\ Arkami-Hamed, S.\ Dimopoulos and G.\ 
Dvali,  {\em New Dimensions at a Millimeter to a Fermi and Superstrings at 
a TeV}, Phys.\ Lett.\ {\bf B436} (1998) 257, [hep-ph/9804398]. 

\bibitem{ST} G.\ Shiu and S.-H.\ Tye, {\em TeV Scale Superstring and Extra 
Dimensions}, Phys.\ Rev.\ {\bf D58} (1998) 106007, [hep-th/9805157].

\bibitem{KT} Z.\ Kakushadze and S.-H.\ Tye, {\em Brane World}, Nucl.\ 
Phys.\ {\bf B548} (1999) 180, [hep-th/9809147].

\bibitem{RS} L.\ Randall and R.\ Sundrum, {\em An Alternative to 
Compactification}, Phys.\ Rev.\ Lett.\ {\bf 83} (1999) 4690, 
[hep-th/9906064]; \ {\em A Large Mass Hierarchy from a 
Small Extra Dimension}, Phys.\ Rev.\ Lett.\ {\bf 83} (1999) 3770, 
[hep-ph/9905221].

\bibitem{SSG} S.\ S.\ Gubser, {\em AdS/CFT and Gravity}, [hep-th/9912001].

\bibitem{JM} J.\ Maldacena, {\em The Large $N$ Limit of Superconformal Field 
Theories and Supergravity}, Adv.\ Theor.\ Math.\ Phys.\ {\bf 2} (1998) 231, 
[hep-th/9711200].

\bibitem{GKP} S.\ S.\ Gubser, I.\ R.\ Klebanov and A.\ M.\ Polyakov, {\em 
Gauge Theory Correlators from Noncritical String Theory}, Phys.\ Lett.\ 
{\bf B428} (1998) 105, [hep-th/9802109].

\bibitem{EW} E.\ Witten, {\em Anti-de Sitter Space and Holography}, Adv.\ 
Math.\ Phys.\ {\bf 2} (1998) 253, [hep-th/9802150].

\bibitem{AH1} M.\ Abou-Zeid and C.\ M.\ Hull, {\em Intrinsic Geometry of 
D-Branes}, 
Phys.\ Lett.\ {\bf B404} (1997) 264, [hep-th/9704021].

\bibitem{AH2} M.Abou-Zeid and C.\ M.\ Hull, {\em Geometric Actions for
D-Branes and M-Branes}, Phys.\ Lett.\ {\bf B428} (1998) 277, 
[hep-th/9802179].

\bibitem{Adler} S.\ L.\ Adler, {\em Einstein Gravity as a Symmetry Breaking 
Effect in Quantum Field Theory}, Rev.\ Mod.\ Phys.\ {\bf 54} (1982) 729.

\bibitem{Zee1} A.\ Zee, {\em A Theory of Gravity Based on the Weyl-Eddington 
Action}, Phys.\ Lett.\ {\bf B109} (1982) 183.

\bibitem{Zee2} A.\ Zee, {\em Einstein Gravity Emerging from Quantum Weyl 
Gravity}, Ann.\ Phys.\ {\bf 151} (1983), 431.

\bibitem{Stelle} K.\ Stelle, {\em Renormalization of Higher Derivative 
Quantum Gravity}, Phys.\ Rev.\ {\bf D16} (1977) 953.

\bibitem{FT} E.\ Fradkin and A.\ Tseytlin, {\em Renormalizable Asymptotically 
Free Quantum Theory of Gravity}, Phys.\ Lett.\ {\bf B104} (1981) 377; \ 
{\em Higher Derivative Quantum Gravity: One Loop Counterterms and Asymptotic 
Freedom}, Nucl.\ Phys.\ {\bf B201} (1982) 469.

\bibitem{Polya} A.\ M.\ Polyakov, {\em Quantum Geometry of Bosonic Strings},
Phys.\ Lett.\ {\bf B103} (1981) 207.

\bibitem{BVH} L.\ Brink, P.\ Di Vecchia and P.\ S.\ Howe, {\em A Locally
Supersymmetric and Reparametrization Invariant Action for the Spinning String},
Phys.\ Lett.\ {\bf B65} (1976) 471.

\bibitem{HT} P.\ S.\ Howe and R.\ W.\ Tucker, {\em A Locally Supersymmetric
and Reparametrization Invariant Action for a Spinning Membrane}, J.\ Phys.\
{\bf A10} (1977) L155.

\bibitem{Deser} S.\ Deser, {\em Scale Invariance and Gravitational Coupling}, 
Ann.\ Phys.\ {\bf 59} (1970) 248.

\bibitem{DG} S.\ Deser and G.\ W.\ Gibbons, {\em Born-Infeld-Einstein 
Actions?}, Class.\ Quant.\ Grav.\ {\bf 15} (1998) L35, [hep-th/9803049].

\bibitem{FTBI} E.\ Fradkin and A.\ Tseytlin, {\em Non-Linear Electrodynamics 
from Quantized 
Strings}, Phys.\ Lett.\ {\bf B163} (1985) 123.
 
\bibitem{RGL} R.\ G.\ Leigh, {\em Dirac-Born Infeld Action from Dirichlet 
Sigma Model},                
Mod.\ Phys.\ Lett.\ {\bf A4} (1989) 2767. 

\bibitem{EWp} E.\ Witten, {\em Bound States of Strings and p-Branes}, 
Nucl.\ Phys.\ {\bf B460} (1996) 335, [hep-th/9510135].                 

\bibitem{CS} C.\ Schmidhuber, {\em D-Brane Actions}, Nucl.\ Phys.\  
{\bf B467} (1996) 146, [hep-th/9601003].                  
           
\bibitem{AE} A.\ Einstein, Sitzungsberichte der Preussische Akademie der
Wissenschaften (1925) 414; \ {\em A Generalization of the Relativistic Theory
of Gravitation}, Ann.\ Math.\ {\bf 46} (1945) 578; \ {\em A Generalized Theory of Gravitation}, Rev.\ Mod.\
Phys.\ {\bf 20} (1948) 35.

\bibitem{JWM1} J.\ W.\ Moffat, {\em Nonsymmetric Gravitational Theory},
J.\ Math.\ Phys.\ {\bf 36} (1995) 3722.

\bibitem{JP} J.\ Polchinski, {\em TASI Lectures on D-Branes}, [hep-th/9611050].

\bibitem{APS} M.\ Aganagic, C.\ Popescu and J.\ H.\ Schwarz, {\em D-brane 
Actions with Local Kappa Symmetry}, Phys.\ Lett.\ {\bf B393} (1997) 311, 
[hep-th/9610249].

\bibitem{CGNSW} M.\ Cederwall, A.\ von Gussich, B.\ E.\ Nilsson, P.\ Sundell 
and A.\ Westerberg, {\em The Dirichlet Super-p-Branes in Ten-Dimensional
Type IIA and IIB Supergravity}, Nucl.\ Phys.\ {\bf B490} (1997) 179, 
[hep-th/9611159].

\bibitem{BTD} E.\ Bergshoeff and P.\ K.\ Townsend, {\em Super D-Branes}, 
Nucl.\ Phys.\ {\bf B490} (1997) 145, [hep-th/9611173].

\bibitem{Arkady} A.\ A.\ Tseytlin, {\em Born-Infeld Action, Supersymmetry 
and String Theory}, [hep-th/9908105].

\bibitem{JHS} M.\ Aganagic, C.\ Popescu and J.\ H.\ Schwarz,
 \ {\em Gauge-Invariant and Gauge-Fixed D-Brane Actions}, Nucl.\ Phys.\ {\bf
B495} (1997) 99,
[hep-th/9612080].

\bibitem{PST} I.\ Bandos, K.\ Lechner, A.\ Nurmagambetov, P.\ Pasti, D.\
Sorokin and M.\ Tonin, {\em Covariant Action for the Super-Five-Brane of
M-Theory}, Phys.\ Rev.\ Lett.\ {\bf 78} (1997) 4332, [hep-th/9701149].

\bibitem{J5} M.\  Aganagic, J.\  Park, C.\  Popescu and J.\  H. Schwarz, {\em
World Volume
Action of the M Theory Five-Brane}, Nucl. Phys. {\bf B496} (1997) 191,
[hep-th/9701166].

\bibitem{CJS} E.\ Cremmer, B.\ Julia and J.\ Scherk, {\em Supergravity 
Theory in Eleven Dimensions}, Phys.\ Lett.\ {\bf B76} (1978) 409.

\bibitem{MD1} M.\ R.\ Douglas, {\em Branes within Branes}, 
[hep-th/9512077].    

\bibitem{GHT} M.\ B.\ Green, C.\ M.\ Hull and P.\ K.\ Townsend, 
{\em                 D-Brane Wess-Zumino  
Actions, T-Duality and the Cosmological Constant},            
Phys.\     Lett.\ {\bf B382}   (1996) 65, [hep-th/9604119].                   


\bibitem{GHM} M.\ B.\ Green, J.\ A.\ Harvey and G.\ Moore, {\em 
I-Brane Inflow and Anomalous Couplings on D-Branes}, Class.\ Quant.\ Grav.\ 
{\bf 14} (1997) 47, [hep-th/9605033].

 
\bibitem{BBG} C.\ Bachas, P.\ Bain and M.\ B.\ Green, {\em Curvature Terms in 
D-Brane Actions and their M-Theory Origin}, {\bf JHEP05} (1999) 011, 
[hep-th/9903210].

\bibitem{JWM2}  J.\ W.\ Moffat,{\em Cosmological Models in the 
Nonsymmetric Gravitational Theory}, [astro-ph/9704300].


\bibitem{FT83} E.\ Fradkin and A.\ Tseytlin, {\em Conformal Anomaly in 
Weyl Theory and Anomaly Free Superconformal Theories}, Phys.\ Lett.\ 
{\bf B134} (1984) 187.


\bibitem{CS99} C.\ Schmidhuber, {\em AdS-Flows and Weyl Gravity}, 
[hep-th/9912155].


\end{thebibliography}
\end{document}